# Feedback-Based Channel Frequency Optimization in Superchannels

Fabiano Locatelli[1], Konstantinos Christodoulopoulos[2]*, Camille Delezoide[3], Josep M. Fàbrega[1], Michela Svaluto Moreolo[1], Laia Nadal[1], Ankush Mahajan[1], Salvatore Spadaro[4]

1: Centre Tecnològic de Telecomunicacions de Catalunya (CTTC)/CERCA, Castelldefels (Barcelona), Spain

2: Nokia Bell Labs, Stuttgart, Germany

3: Nokia Bell Labs, Paris, France

3: Universitat Politècnica de Catalunya, Barcelona, Spain

*Corresponding author: konstantinos.1.christodoulopoulos@nokia-bell-labs.com

**Abstract**

Superchannels leverage the flexibility of elastic optical networks and pave the way to higher capacity channels in space division multiplexing (SDM) networks. A superchannel consists of subchannels to which continuous spectral grid slots are assigned. To guarantee superchannel operation, we need to account for soft failures, e.g., laser drifts causing interference between subchannels, wavelength-dependent performance variations, and filter misalignments affecting the edge subchannels. This is achieved by reserving spectral guardband between subchannels or by employing a lower modulation format. We propose a process that dynamically retunes the subchannel transmitter (TX) lasers to compensate for soft failures during operation and optimizes the total capacity or the minimum subchannel quality of transmission (QoT) performance. We use an iterative stochastic subgradient method that at each iteration probes the network and leverages monitoring information, particularly subchannels signal-to-noise ratio (SNR) values, to optimize the TX frequencies. Our results indicate that our proposed method always approaches the optima found with an exhaustive search technique, unsuitable for operating networks, irrespective of the subchannel number, modulation format, roll-off factor, filters bandwidth, and starting frequencies. Considering a four-subchannel superchannel, the proposed method achieves 2.47 dB and 3.73 dB improvements for a typical soft failure of ±2 GHz subchannel frequency drifts around the optimum, for the two examined objectives.

## 1. Introduction

During the past few years, in addition to the dense wavelength division multiplexing (DWDM) paradigm, in which narrow spacing between fixed channels was considered [1], a new flexible DWDM approach was defined [2]. This new paradigm, also known as flex-grid DWDM or elastic optical networks (EON), envisioned smaller frequency slots (central frequency granularity of 6.25 GHz and a slot width of 12.5 GHz) that could be combined to form channels of variable widths [3]. Flex-grid DWDM enabled the transmission of heterogeneous channels with high spectral efficiency over the same network.

Reconfigurable optical add/drop multiplexer (ROADM) nodes include wavelength selective switches (WSSs) to allow the individual switching of optical channels [4]. The WSS behaves essentially as an optical filter, which filters and selects the channels from the ingress port(s) to forward to the egress port(s) [5]. Thus, in a ROADM-based optical network, the spectral densification enabled by the flex-grid paradigm also implies higher filtering penalties for the channels. Apart from the filtering process itself, these penalties come from the well-known filter-related impairments, i.e., filter cascade effect (FCE), also known as filter narrowing effect, and transmitter (TX)-filter frequency detuning [6]. A possible solution to overcome such issues is represented by superchannels. The main purpose of superchannels is to achieve high rates with limited electronics, but they can also reduce filtering effects and achieve higher spectral efficiency. A superchannel includes multiple channels (or subchannels, which is the term used in the rest of this paper) allocated into a contiguous set of spectral slots with a narrow guard band in between them [7]. Fig. 1 depicts a superchannel with $N$ subchannels. When travelling through the optical network nodes (i.e., crossing the ROADMs and the WSSs composing them), the superchannel behaves as a single entity [8]. So, the WSS filters at the ROADM nodes are set to switch all the superchannel related slots, a configuration that is also referred to as superfilter. This allows reducing the inter-channel distances, therefore improving the total spectral efficiency. Superchannels can be created with multi-laser TX or single-laser TX that form the comb electrically. We propose a solution that applies to multi-laser TX; although the market is today not moving in this direction due to a high cost, multi-laser TXs could appear in the future. Moreover, extension of our solution can be applied to optimizing the frequencies of all the channels in the network, a solution that is applicable to any type of network with single- or multi-laser and single- or multi-carrier TXs.

Nowadays, space division multiplexing (SDM) is attracting much attention, and it is considered one of the most promising approaches to cope with the ever-increasing capacity requests [9]. In SDM-based networks, either multifiber links or multicore / multimode fibers are employed that require the evolution of the single-channel granularity of the ROADM-based networks towards a bulkier approach. In this context, superchannels, which are the next level of switching granularity, are expected to be extensive used [10]. Thus, the proposed solution, although evaluated here for single-core and single-fiber networks, would find good applicability in an SDM network.

A common metric to describe the superchannel spectral efficiency class is the ratio of the subchannel distance (or allocated bandwidth) to the symbol rate [11]. If such ratio is between 1 and 1.2 (e.g., a 32 GBd signal with a subchannel spacing of 37.5 GHz), then the class is defined as quasi-Nyquist-WDM, whereas if the distance equals the symbol rate (i.e., the ratio is equal to 1), the class is defined as Nyquist-WDM. Finally, if the distance is lower than the symbol rate (i.e., a ratio less than 1), the class is defined as super-Nyquist-WDM. It is worth noting that such classification applies to superchannels with equally spaced subchannels (i.e., equidistant superchannels) and where all subchannels have the same symbol rate and the same roll-off factor (i.e., uniform superchannels) [11].

When taking into account superchannels, we have to consider their main limitations, which are interference and filtering penalties. In particular, interference comes in two forms, cross-phase modulation (XPM) also referred to as cross-channel nonlinear interference (XCI) in the GN model [12], occurring among all subchannels (but mainly between close adjacent ones) and (linear) adjacent channel interference, or crosstalk, occurring only between adjacent ones. The latter, which is the dominant effect, can be caused by uncontrolled drifts of the TX lasers (e.g., as a consequence of ageing) and translates into a degradation of the subchannel quality of transmission (QoT), such as the signal-to-noise ratio (SNR) value [13]. Instead, the filtering penalties only affect the two external subchannels (i.e., $f_1$ and $f_N$ in Fig. 1). These, if combined with the potential misalignment between filters and TXs, result in SNR degradation for the two external subchannels. Finally, we must consider the wavelength-dependent effects (e.g. amplifier gain ripples), resulting in differential degradation of the subchannel SNR values. To this end, the worst superchannel performance should be accounted for. To compensate for all such effects, we can operate the superchannel at lower modulation formats and/or use spectral guardband (higher inter-channel distance), sacrificing capacity and/or spectral efficiency. Note that these effects cannot be known before the superchannel is established and operates, while they might also vary with time. Therefore, ways to monitor and eventually correct such behaviors while the network operates are needed.

To overcome filter related impairments for a single channel while the network operates, in [14], the authors presented a method for the orchestrated alignment of a single channel with the cascade of optical filters. The solution relied on a control plane that moved the channel TX frequency to maximize (or minimize) a specifically chosen metric. The effect of each correction was then measured from the signal power spectral density (PSD), and based on that, the control plane took the following decisions. This control loop process ended once it stabilized at a specific TX frequency value. A similar approach was proposed in [15] and [16], where a field demonstration of an autonomic, self-reconfigurable network was presented. In particular, the authors leveraged an automated control loop to mitigate distortions caused by TX laser and filter misalignments. To do so, the software-defined networking (SDN) controller triggered a specific step adjustment of the TX frequency by monitoring, storing, and processing one or more QoT metrics, such as the bit error ratio (BER) or the SNR [17].

In this paper, we extend these previous works by proposing a method to optimize the frequencies of the superchannel subchannels. Similar issues are also faced at the network level, where adjacent channels that enter a ROADM at the same ingress port and are forwarded to the same egress port share a superfilter configuration. The misalignment of the channels sharing such superfilter would create interference similar to that in superchannels. So, the proposed method can be eventually extended to reduce that interference and align the whole network channels; such a scenario is left for future work.

The traditional way to design a superchannel is to leverage a physical layer model (PLM) and define the TX and receiver (RX) configurations, including subchannel distances and superfilter configuration. Typically, this is done for equidistant and uniform configurations, as in [8]. However, such a strategy would be optimal in ideal systems, whereas in reality, several limitations such as the laser drifts, filter misalignments, and wavelength-dependent effects outlined above should also be considered. These result in variations of the subchannels QoT (e.g., their SNR values) and make equidistant superchannels suboptimal and unattainable, therefore requiring the use of margins. Thus, using an approach that leverages the network feedback and takes into account its current conditions, such as the one presented in this paper, can sense and correct the above issues, trace the real optimum, reduce the employed margin, and increase the superchannel efficiency. Note that superchannels with time-frequency packing [18] compensate interchannel interference in DSP; the proposed method could work in parallel and/or simplify the algorithms and the hardware complexity of such TRXs.

We formulate the problem of optimizing the superchannel frequencies as a nonlinear optimization one. Our objective is to identify the set of subchannel frequencies that optimizes a function of their QoT. To be more specific, we consider the maximization of the average of all the subchannel SNR values, corresponding to the total capacity maximization, and the minimum subchannel SNR value, corresponding, for instance, to make all the subchannels feasible for a specific modulation format. The nonlinear dependence comes from the interference (with two components: XPM and linear crosstalk) and the filtering penalties for the edge subchannels. The SNR of a subchannel decreases nonlinearly as the subchannel moves closer to its adjacent ones or the filter edges. In particular, in Section 2.1, we report the results of simulations showing the channel SNR function to be concave around the typical operation point (equal distance). A function is concave when it has an inverted bowl shape, that is, if we choose any two points on its graph and draw a line segment to join them, such segment will always entirely lie below the graph. A concave function can be easily transformed to convex by taking its negative. As so, we can apply convex optimization techniques to solve the considered problem. However, the convex function might not be very smooth due to random noise, TX/RX imperfections, and monitoring errors. The method we present uses a stochastic subgradient, an algorithm that can solve convex problems with zero mean noise [19] and several other nonconvex problem classes [20], e.g., quasi-convex problems. So, we chose the proposed algorithm to be robust and able to optimize under the uncertainties that are expected to appear in a real network.

More specifically, we propose an iterative closed control loop process to optimize the superchannel frequencies based on the stochastic subgradient algorithm. The proposed automatized optimization process uses monitoring information of the subchannels (i.e., their SNR values monitored at their RXs). It probes the superchannel with new frequencies, monitors the outcomes, and moves to new frequencies (note that the stochastic subgradient might not move at each step towards the optimum). The novelty of this work lies on the use of monitoring information in a closed-loop optimization process that is robust to monitoring or other random errors. Previous works base their optimization on closed-equation physical layer models that avoid certain impairments (e.g. wavelength dependent loses, ASE noise randomization) and/or are not robust against uncertainties or real time network conditions.

We evaluate the considered solution integrating the proposed optimization algorithm within the VPIphotonics simulation tool [21], based on split step Fourier transformation [22]. To maximize the desired objective function, we feed the data monitored/collected from VPI to the optimization algorithm. Then, using the algorithm outputs, we adjust the subchannel frequencies accordingly and again monitor the output of VPI and pass it to the algorithm. In turn, a new algorithm iteration starts, and the closed control loop is realized as if the superchannel was adjusted in an operating network. We believe this approach to be universal and applicable in different contexts, regardless of the considered ROADM architecture, network disaggregation level or specific segment (e.g., metro, core, etc.). The results obtained show that independently from the considered configuration (i.e., modulation format, roll-off factor, filter bandwidth, span length, number of filters, and number of subchannels), the proposed solution can improve the superchannel average SNR and the minimum subchannel SNR.

The rest of the paper is organized as follows. In Section 2, we formulate the optimization problem we are aiming to solve. In Section 3, we describe the solution proposed to address the considered optimization problem. In Section 4, we present the results obtained using the proposed approach. Finally, Section 5 concludes the paper.

## 2. Problem Formulation

Throughout this paper, we will denote scalars with italic and vectors with bold font. We assume a superchannel with an allocated spectrum $BW_{SC}$ centered around the frequency $f_{SC}$ and composed by a set of $N$ subchannels, as depicted in Fig. 1. In addition, following the definition we introduced in Section 1, we initially assume the superchannel to be equidistant and uniform. We consider the spectral shaping of the subchannels to be defined by a square root raised cosine (RRC) with a roll-off factor equal to $a$. If we assume a single channel with symbol rate $R_S$ and roll-off factor $a$ to pass through an optical filter with bandwidth $BW$, the minimum required filter bandwidth $BW_{min}$, for the channel to not heavily suffer from filtering effects, would be $BW_{min} = R_S (1 + a)$. For instance, considering a single channel with 32 GBd symbol rate and 0.1 roll-off factor that crosses an optical filter, the minimum required filter bandwidth to avoid filtering-related penalties would be 35.2 GHz. As so, considering filters with 3 dB bandwidths equal to 37.5 GHz (i.e., following the flex-grid definition) would result in enough spectral guardband space, so low penalty, for such transmission.

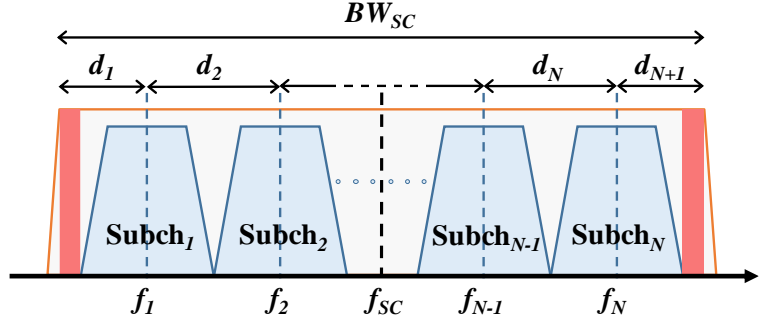

Fig. 1. Example of a superchannel with *N* subchannels. The two red bands at the superchannel sides represent the guard bands.

Similarly, if we consider a superchannel, the minimum subchannel distance to have crosstalk-free (or very low crosstalk) interactions with the adjacent subchannels can be defined as $d_{min} = R_S (1 + a)$. For instance, in a superchannel with subchannels defined by 32 GBd symbol rate and 0.1 roll-off factor, when the distance between subchannels is larger than 35.2 GHz, the channels will have very low crosstalk. Note that an underlying assumption about the subchannels being equidistant and uniform was made in the above calculation. This definition can be extended to cover nonequidistant and nonuniform superchannels. In addition, we also assume the superchannel to be established in a DWDM/flex-grid optical network and that it crosses at least one filter with 3 dB bandwidth $BW_{SC}$ and central frequency $f_{SC}$ (which we also refer to as a superfilter).

We will now extend the notation to cover nonequidistant superchannels. We denote by **D** = [$d_1$, $d_2$, ..., $d_{N+1}$] the vector of length *N*+1, where $d_1$ and $d_{N+1}$ represent the distances between the superchannel filter sides (3 dB) and the first and last superchannel subchannel central frequencies, $f_1$ and $f_{N+1}$, respectively. Moreover, $d_n$, with $2 \leq n \leq N$, represents the generic distance between the (*n*-1)-th subchannel central frequency, $f_{n-1}$, and the *n*-th subchannel central frequency, $f_n$. Note that the proposed solution is mathematically described using the distance vector **D**, but in text, it is described as the optimization of frequencies; the translation from one to another is straightforward.

The quality of transmission (QoT), as indicated by a generic metric that includes all impairments e.g. the SNR, EVM, or BER, values of the superchannel subchannels are directly correlated to the distances between their central frequencies. In the following we will formulate the problem using the SNR as the QoT metric of interest, but the problem formulation and the proposed solution hold for other generic QoT metrics, such as EVM and BER. We denote with $SNR_n$(**D**) the SNR value of the *n*-th subchannel. In particular, $SNR_n$(**D**) mainly depends on distances $d_n$ and $d_{n+1}$, which are those separating subchannel *n* from its two adjacent ones, since when their spectra overlap, (linear) crosstalk is created, and the highest XPM also occurs. Moreover, because the XPM caused by all subchannels gives a small contribution, $SNR_n$ also depends on all subchannel distances, and thus on **D**. For instance, bringing one subchannel closer to its left adjacent would increase their crosstalk and interference, resulting in a reduction of their SNR values with respect to their distance. On the other hand, such a shift would reduce the crosstalk and interference of the moved subchannel with its right adjacent one. We will see in Section 2.1, that the SNR function is concave around the point corresponding to the equal distance of a subchannel from its adjacent ones.

The optimization problem needs to identify the proper distance set **D*** that maximizes specific functions of the subchannel SNR values. In particular, we consider two optimization functions:

i) the maximization of the superchannel average SNR value, equal to the sum of SNR values of all subchannels divided by their number *N*, and

ii) the maximization of the minimum subchannel SNR value.

The former objective (Obj#1) is directly related to the maximization of the superchannel total capacity, whereas the latter (Obj#2) targets to make all subchannels feasible, with respect to a specific SNR threshold for the used modulation format. We can formulate the optimization problem (with the two objectives mentioned above) as follows:

$$max(h(\mathbf{D})), \quad (1)$$

Where

$$h(\mathbf{D}) = \frac{\sum_{n=1}^{N} SNR_n(\mathbf{D})}{N}, \quad (2)$$

or

$$h(\mathbf{D}) = min_n(SNR_n(\mathbf{D})). \quad (3)$$

In addition, the following set of constraints applies:

$$\sum_{n=1}^{N+1} d_n = BW_{SC} \quad n = 1, \dots, N, \quad (4)$$

$$d_n = \begin{cases} \frac{R_S}{4}, & n = 1, N+1 \\ \frac{R_S}{2}, & 1 < n < N+1 \end{cases}. \quad (5)$$

Constraint (4) keeps the subchannel distances within the filter bandwidth. Constraint (5) put a lower limit on the distances between adjacent subchannels and outer subchannels and filter, based on the subchannel symbol rate, but it is optional and can be skipped. Note that we assumed the SNR as the QoT metric in the above definition, but we could convert it to a corresponding SNR margin by considering the SNR threshold of the used modulation format. We can also define the problem with other generic QoT metrics, such as the BER. Also, note that in the above formulation, we assumed a uniform superchannel where all subchannels have the same symbol rate, the same roll-off factor, and the same modulation format. However, the model and the proposed method can be extended to nonuniform cases.

**2.1. SUBCHANNEL DISTANCE AND SNR FUNCTION PROPERTIES**

To understand the relation between the subchannel distances and their SNR values, we implemented in VPI a simulation setup considering three channels with nonidentical pseudo-random binary sequences (PRBSs). We generated three 32 GBd quadrature phase shift keying (QPSK) modulated signals, pulse shaped with an RRC filter with 0.1 roll-off factor. In Fig. 2, we plot the SNR values of the second channel as a function of its distance $d_1$ from the first channel, while its distance $d_2$ from the third channel was set to keep $d_{1,2} = d_1 + d_2 = 69$ GHz. Stated differently, we kept the first and third channel central frequencies constant and shifted the second channel. We plot the SNR values for a back-to-back (B2B) scenario and transmission over two and ten standard single-mode fiber (SSMF) spans of 80 km length, with 0.2 dB/km attenuation coefficient and 16.7 ps/nm/km dispersion coefficient. Each SSMF span was followed by an erbium-doped fiber amplifier (EDFA) with a 5.5 dB noise figure (NF) and gain that matched the span loss.

We observe in Fig. 2 that the SNR of the intermediate (second) channel is a concave function of $d_1$ around the equidistant point ($d_1 = d_2 \pm 2$ GHz). As so, the considered objective functions, which are combinations of such functions, are also concave. Varying the channel distance, the resulting change in the XPM is relatively small, so the dominant effect observed in Fig. 2 is the linear crosstalk between adjacent channels. Thus, the shape of the SNR function depends on the crosstalk, which in turn depends on the channel shaping. In particular, the crosstalk contribution from the adjacent channel is convex: as we move the second channel towards the first, the amount of area in which the signals overlap increases. The crosstalk is proportional to the overlapping area integral. As the channels distance reduce the overlap increases, the integral of an increasing function is convex, and thus crosstalk is convex. Similar observations would hold if we moved to the other side. The sum of two convex functions is convex, so the total crosstalk contribution is convex. Finally, the SNR depends inversely on the crosstalk and thus is concave. Note that the above hold for any generic QoT metric: EVM is concave similar to the SNR, while the BER is convex, as also shown in Fig. 2 (VPI simulations). Note that the proposed solution works for any generic QoT metric if it is a concave or convex function of the channel distances.

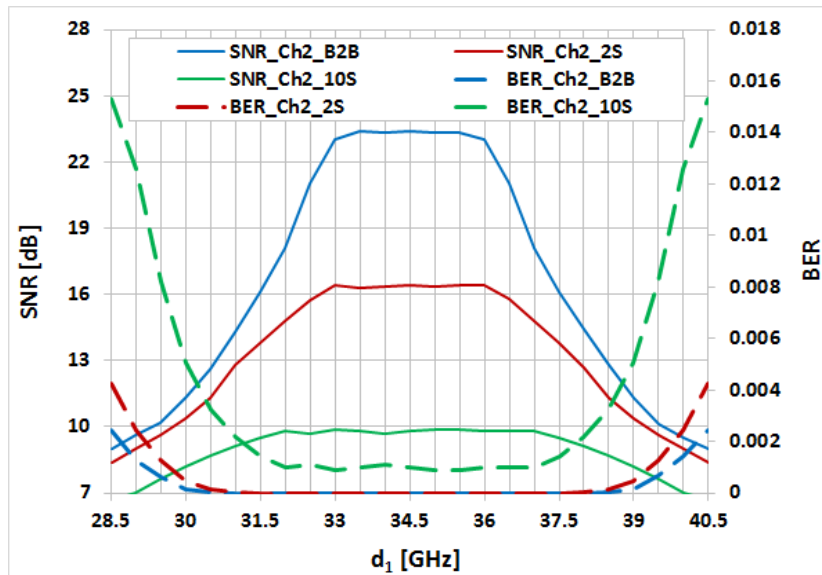

Fig. 1. Relation between the SNR and BER value of the second channel and its distance with the first one, for B2B, two SSMF spans (160 km) and ten SSMF spans (800 km) scenarios.

Based on the above, we expect a single or a continuous set of distances $d_1$ that exhibits the maximum SNR value (or close to maximum as discussed in the following) for the second channel, making gradient-based methods suitable to solve this problem. Note that a similar shape appears when one side of the channel faces the filter. Regarding the number of SSMF spans, we observe that transmitting the signals over the spans reduces the reference SNR value and flattens the shape of the SNR curve. Still, the concave shape is maintained, and the proposed optimization process holds, as indicated by our results. The causes of the SNR curve flattening are the noise generated by the EDFAs, which accumulates between the channels, and the widening of the channel spectra caused by the dispersion. Thus, the channel spectra are shallower and their sides less sharp, resulting in less crosstalk area when they overlap and flatter SNR curves.

To be even more specific, although from Fig. 2 it might seem that there is a set of distances $d_1$ for which the maximum SNR value flattens out, if we zoom in that area, we would see minor SNR variations caused by random noise, wavelength-dependent penalties, and other factors. The increasing/decreasing sides are also not smooth. Furthermore, in real systems, we would also have additional effects coming from TX/RX imperfections and monitoring error (note that we assume here to monitor the SNR or the pre-FEC BER values at the coherent RXs). Therefore, there are several local maxima close to the global in the flat area, and maybe at the sides. Even more, the global and local optima would vary as a function of time due to the short- and medium-term impairments. For these reasons, we decided to use the stochastic subgradient method with a fixed step length, which is robust, can optimize in presence of noise and uncertainties, and works even with nonlinear and nonconcave (or nonconvex) problems.

Thus, in our optimization, we target to reach the flat area and any local maxima close to the global one. Our results showed that the obtained solution was consistently within 0.2 dB from the optimum, using the proposed optimization method based on a stochastic subgradient.

**2.2. Dynamic Superchannel Optimization**

As also discussed in the introduction, typically, a superchannel would be designed with appropriate TX and RX configuration and equal distances between subchannels [8]. However, there are two main reasons that equal distance will not be the optimal in a real network. Firstly, imperfections such as those related to the subchannel TXs and RXs, dynamic impairments, wavelength-dependent losses / performance variations (e.g., amplifier gain ripple), and ageing of the various network elements make unequal the QoT of even two identical channels. Such factors result in subchannel QoT variations (e.g., their SNR values) and make equidistant superchannels suboptimal. Secondly, due to laser drifts, the control plane would configure the superchannel as equidistant, but that would be unattainable. The second issue could be corrected by a closed control loop that monitors the received subchannel frequencies, but the first could not.

For the above reasons, margins in QoT or additional spectra between the subchannels are typically used. Note that such effects are unknown before the superchannel is established and operates, and they might also vary with time. For example, assuming a ±2 GHz drift for the second channel of Fig. 2 in the two spans case, we end up with a 0.7 dB lower SNR. To compensate for this, this dB amount should be considered as a margin when selecting the superchannel modulation format (actually, the margin should be higher because we should consider the worst case, in which the first and third channels also drift).

To reduce the margin or increase the efficiency of the superchannel, we need a feedback-based approach that dynamically interacts with the superchannel and the network, understands the current conditions, and corrects and compensates these effects. In the following section, we propose such approach that dynamically optimizes the superchannel as the network operates. The propose method could run periodically or on demand when e.g. through monitoring we observe that the performance (SNR or BER) or a superchannel has fallen below a specified threshold.

**3. Closed control loop and stochastic subgradient method**

The proposed solution implements a closed control loop and employs an iterative optimization algorithm based on the stochastic subgradient method with a fixed step length. We assume the optimization algorithm to reside in the central SDN controller and that the SDN controller has the appropriate interfaces to configure the TX lasers and monitor the SNR (or the pre-FEC BER) values of the subchannels. The optimization algorithm employs a subroutine to specify the probes, the configurations enforced to the network through the control plane. Hence, the algorithm receives the monitored data (SNR values), which are the outcomes of the applied configuration, to identify the information that needs for an intermediate optimization step. This process is repeated at each iteration. We also refer to this technique as optimization with monitoring probes. A representation of such a scheme is shown in Fig. 3.

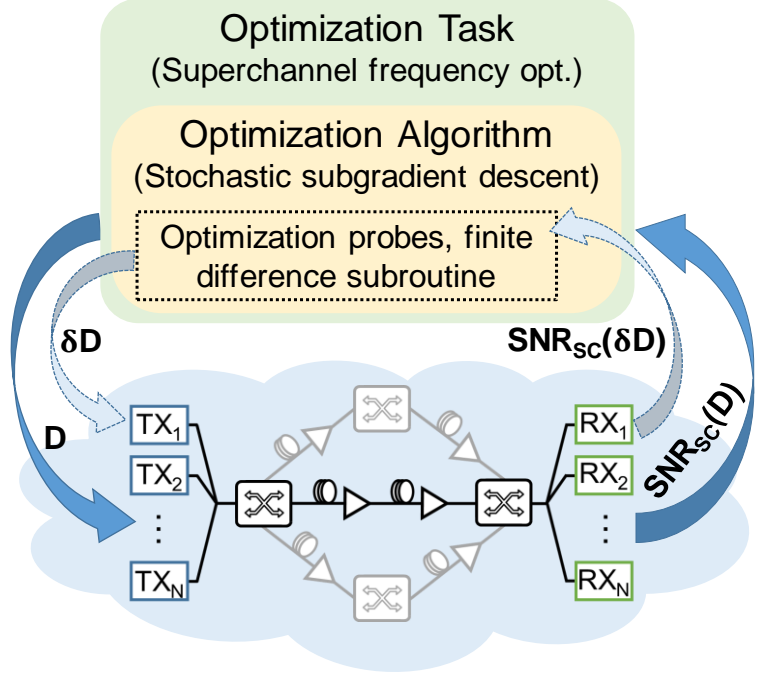

Fig. 2. Schematic representation of the implemented optimization loop with monitoring probes.

To be more specific, we address the frequency optimization of the subchannels with a typical objective, such as the maximization of the subchannel average SNR value or the subchannel minimum SNR value, as discussed in Section 2. This problem is concave with respect to the distances of the subchannels, as discussed in Section 2.1. Hence, considering the negative of the SNR function, we can easily transform the problem into the minimization of a convex function, for which several known optimization techniques exist. Optimization algorithms such as (sub)gradient method, interior point, trust-region-reflective, etc., are iterative, meaning that they need to calculate (first or second order) partial derivatives at each iteration. In particular, we choose to use a method based on a stochastic subgradient that can find the optimum of a noisy convex problem (or, in the more generic form, a nonlinear problem with the specific properties discussed in Section 2.1) in a polynomial number of steps. For optimization problems involving the optical physical layer and the capability to probe the network (configure and monitor the outcome), a typical way to find the partial derivatives is by using a subroutine that implements the finite difference method [23].

To calculate the (sub)gradient of the objective function h, we need to configure new frequencies and monitor the related changes in the subchannel SNR values. Specifically, following the notation we previously introduced, we assume a superchannel consisting of a set of *N* subchannels and a distance vector **D**. We denote by $\boldsymbol{\delta D_n}$ the distance vector for which the central frequency of the *n*-th subchannel has increased by $f_{step}$, which we refer to as the frequency probe step, and in the optimization process is considered to be a fixed step length. This implies that $f_{step}$ is subtracted from $d_{n-1}$ and added to $d_n$.

We denote the SNR of the shifted subchannel n with $SNR_n(\boldsymbol{\delta D_n})$, whereas the SNR of all subchannels with $SNR_{SC}(\boldsymbol{\delta D_n}) = [SNR_1(\boldsymbol{\delta D_n}), ..., SNR_N(\boldsymbol{\delta D_n})]$. Note that the change in the central frequency of the single subchannel n mainly affects the SNR values of the adjacent subchannels (*n*-1 and *n*+1) through (linear) crosstalk, but more lightly also of all the other subchannels through XPM. The partial derivative $g_n$ of the optimization objective function h for the *n*-th subchannel is given by

$$g_n = \frac{h(\boldsymbol{D}) - h(\boldsymbol{\delta D_n})}{f_{step}}. \qquad (6)$$

Depending upon the objective function $h$, this involves specific operations with vectors $SNR_{SC}(\mathbf{D})$ and $SNR_{SC}(\mathbf{\delta D_n})$, as indicated in the two considered objectives (2) and (3). Thus, to calculate the (sub)gradient with the finite difference method, we need to probe with $\mathbf{\delta D_n}$ and to monitor $SNR_{SC}(\mathbf{\delta D_n})$. We repeat this for the subchannels that are chosen (stochastically) for each algorithm iteration.

The algorithm at each iteration calculates the subgradients of a set of randomly selected subchannels through the finite difference subroutine. In particular, we use the minibatch option of the stochastic method, in which at each iteration, we randomly choose a set of $M$ subchannels to probe and monitor. The stochastic subgradient method with a fixed step length $f_{step}$ and monitoring probes runs $I$ iterations to converge or approach the optimum, stopping when the calculated objective improvement rate falls below a specified threshold. In particular, if $f_{step} = 2\varepsilon/G$, where $G$ represents a bound on the gradient such as $|h(\mathbf{D_u}) - h(\mathbf{D_v})| < G\|\mathbf{D_u} - \mathbf{D_v}\|_2$, valid for any $\mathbf{D_u}$ and $\mathbf{D_v}$, then the subgradient method converges to the optimum within $\varepsilon$. Specifically

$$h_{best}^i - h^* \leq \varepsilon, \qquad (7)$$

where, $h^i_{best}$ and $h^*$ represent the objective functions related to the $i$-th iteration and the optimal solution, respectively. Therefore, the required number of steps to converge towards the optimum within $\varepsilon$ is given by $(RG/\varepsilon)^2$, where $R$ is the distance of the starting point from the optimum, such as $R \geq \|\mathbf{D}^{(1)} - \mathbf{D}^*\|_2$. Note that the subgradient method discussed above is a batch method; it calculates the subgradients for all the functions summed in the objective (e.g., the SNR function of the different subchannels). The stochastic subgradient method that we adopt has a bound on the convergence rate given by $O(N^2/\varepsilon^2)$. When we use minibatches of size $M$, we have a bound on the convergence rate given by $O(N^2/M^2\varepsilon^2)$ [24]. So, the bound on the expected number of iterations of the stochastic subgradient method with minibatches is given by

$$I = \left(\frac{NRG}{M\varepsilon}\right)^2. \qquad (8)$$

In turn, each iteration is composed of $M$ calls of the (sub)gradient identification subroutine, corresponding to the times the algorithm probes the network and monitors the SNR (or pre-FEC BER) values. We denote by $t_{mon}$ the monitoring time needed to monitor all the $N$ subchannels simultaneously. Monitoring is typically done every 15 minutes in today networks, but the new generation of telemetry-based solutions can speed up this process to subsecond timescales [25]. However, the monitoring time here is constrained by other factors. Once a reconfiguration is decided and applied, the time needed by the transponders to adapt to such change and by the network to reach a stable state (e.g., for transient effects to settle) must also be considered. Moreover, monitoring the SNR (or pre-FEC BER) is not instantaneous but requires to average over a specific period. It stands to reason that the monitoring accuracy increases with time. If monitoring is fast and the error is high, the algorithm will have to perform more iterations. Thus, although we will have a small $t_{mon}$, we will pay that with a higher iteration number $I$. Note that the stochastic subgradient method finds the optimum in a polynomial number of iterations $I$. Based on the above, we expect $t_{mon}$ to range from tens of seconds to minutes, depending upon the network complexity, the monitoring plane, the targeted monitoring error, and other factors [24]. However, once the monitoring information is forwarded to the algorithm, the time $t_{calc}$ that the algorithm needs to calculate the (sub)gradients and also the following frequencies is relatively low (within seconds) compared to the monitoring time (i.e., $t_{mon} \gg t_{calc}$). So, with the proposed approach, under the assumption that the $N$ subchannels are monitored in parallel, the total optimization time $T_{opt}$ is given by

$$T_{opt} = I \cdot (M \cdot t_{mon} + t_{calc}) \approx I \cdot M \cdot t_{mon}. \qquad (9)$$

Returning to the optimization problem, the set of distances **D** is defined by the central frequencies of the subchannel TX lasers. However, the lasers cannot be configured with infinite accuracy. Moreover, since we rely on physical layer monitoring information and because of physical layer variations (mainly short and medium terms affect the optimization) and monitoring errors, we are not able to observe fine differences in the monitored SNR values. For these reasons, as discussed above, we use for the gradient identification in both the optimization problems, a frequency step of $f_{step}$. For example, such a step can be equal to 0.25 GHz, 0.5 GHz or 1 GHz. The value of $f_{step}$ should be chosen according to the laser configuration capabilities and the uncertainties coming from variability and monitoring errors. In particular, smaller values tend to make the stochastic subgradient method slower but give better granularity, and therefore the optima can be approached more precisely. Compared to the optima found with a brute-force approach, our results indicate good convergence performance for 0.25 GHz and 0.5 GHz frequency steps, whereas the performance tends slightly to worsen for a frequency step of 1 GHz.

Note that the monitoring probes used to identify the (sub)gradient are not a universal solution. A monitoring probe here refers to the configuration of new frequencies for one or more subchannel lasers and monitoring all the subchannel SNR values at their RXs. Each optimization problem requires its specific definition of monitoring probes. For some problems, monitoring probes might not be available, e.g., problems involving establishing or releasing connections. Moreover, the monitoring probes approach involves several interactions with the network, which might be time-consuming. Therefore, alternative methods that use a PLM /a QoT estimator such as the GN model [12] are applicable and might be preferable according to the considered problem [26]. Note that the proposed approach based on monitoring probes achieves the optimal performance as long as the problem has specific properties (convexity or unimodularity) and an appropriate algorithm is used. However, such a solution might take a long time due to the delay introduced by the monitoring process. On the other hand, using a PLM is much faster and without monitoring noise, but its inaccuracies might mislead the optimization. Such concepts are studied in [26] for another optimization problem and are left for future investigations for the problem considered in this paper.

Regarding the optimization algorithm, we adapted the stochastic subgradient algorithm with minibatches. Given the subgradients of *M* subchannels, specified with the monitoring probes / finite difference method mentioned above, the algorithm decides on the next move and identifies the new central frequencies of all the subchannels. Note that the algorithm might decide to make a more complex action (move several subchannels) than the simple movements performed in the gradient identification subroutine. The algorithm output is mapped to the network, and a new probing and monitoring phase starts. The stochastic subgradient algorithm converges to the optimum, with a specific accuracy, in a polynomial number of iterations [24]. However, although the algorithm selects/calculates new variable values to improve the objective at each step, such choices might not be effective when applied to the actual system because of the noise. So, the best result is memorized and reused as starting point in case the objective falls. Once the algorithm converges (no further improvement of the objective is calculated for a certain amount of iterations, with a defined accuracy set, for instance, to 0.2 dB), the process terminates. The configuration saved as the best one up to that point corresponds to the optimized frequency set for the superchannel subchannels.

## 4. Results and discussions

To quantify the benefits of the proposed superchannel frequency optimization method, we carried out simulations using VPI and MATLAB. In particular, we co-simulated the superchannel and its transmission in VPI, whereas the finite differences subroutine and the stochastic subgradient algorithm were developed in MATLAB. The specific physical layer model (VPI) is based on split step Fourier transformation [22] and was chosen since it models the key physical layer impairments that affect the proposed solution. In particular, it includes TX laser linewidth, nonidentical PRBS and pulse shaping with different roll-offs at TX side. It also models linear crosstalk, wavelength dependent deteriorations such detailed nonlinear interference, EDFA gain ripple, and ASE noise randomization. These create the QoT fluctuations observed in Fig. 2 which are hard to optimize against. Without such effects, simpler optimization methods (e.g. gradient descent) or even offline calculations would work, but could fail in reality. Also note that by using such physical model, we were constrained to perform simulations for a single superchannel and few links/nodes. However, we argue that using a bigger and/or loaded network will not affect the proposed method nor change the results and conclusions of the paper. In a loaded network the addition connections will create additional XPM and other nonlinear effects but these will remain the same, or have very small variations, over each monitoring/optimization step. Constants do not affect optimization; in addition the proposed method is robust against small variations.

To be more specific, we implemented in MATLAB the frequency optimization algorithm to maximize: the superchannel average SNR value, corresponding to the maximization of the superchannel total capacity (Obj#1), or the lowest subchannel SNR value, corresponding to, for instance, making all subchannels feasible for a specific modulation format (Obj#2), as discussed in Section 2. In the co-simulation setting, VPI takes as input the subchannel frequencies, performs the detailed transmission simulations and calculates the SNR values of the received subchannels. The MATLAB code implements the finite difference subroutine; it identifies which frequencies **δD** to probe the superchannel, passes those to VPI, and finally receives the monitored SNR values $SNR_{SC}(\mathbf{\delta D_n})$. After a specific number of such probes (minibatch of $M = 2$ subchannels in our simulations), the optimization algorithm identifies their subgradients and uses them to calculate the following frequencies. This process is repeated until the algorithm converges to the optimal solution, which without considering the monitoring phase, we measured to require a time in the order of seconds. In addition, we kept track of the best-achieved solution and then stopped the algorithm once the estimated improvement was below a specific threshold (the optimality tolerance in the following results was set to 0.2 dB). Note that, as discussed earlier, the same method is applicable when monitoring the pre-FEC BER values of the sub-channels.

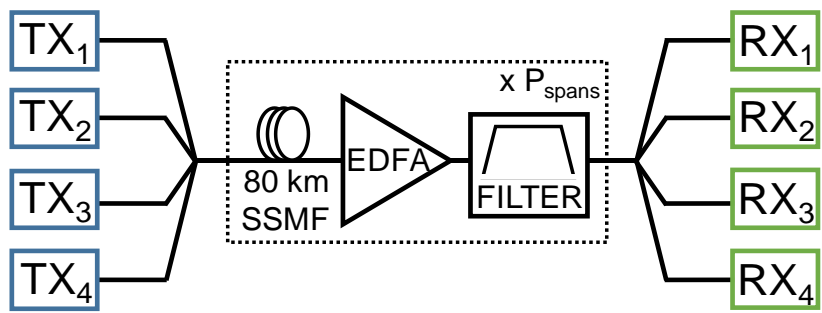

Fig. 3. The VPIphotonics simulation setup. TX: transmitter, SSMF: standard single-mode fiber, EDFA: erbium-doped fiber amplifier, RX: receiver.

To simulate a superchannel with four subchannels, we implemented in VPIphotonics the setup depicted in Fig. 4. We generated four 32 GBd QPSK modulated signals that formed four quasi-Nyquist-WDM subchannels symmetrically centered around 193.1 THz. For their pulse shaping, we employed an RRC filter with roll-off factors equal to 0.1 and 0.15. We set the launch power of each TX laser at 0 dBm. As starting condition, we always considered the four signals equally spaced by 34.5 GHz, resulting in quasi-Nyquist-WDM transmissions (i.e., the ratio between inter-channel distance and symbol rate between 1 and 1.2). In addition to the equally distanced configuration, we also considered ten random starting frequencies, which were ±2 GHz shifted from the equally distanced setting. This was done in order to emulate soft failure scenarios. Then, we multiplexed the four subchannels and transmitted them in a loop composed by the cascade of an 80 km length SSMF, with 0.2 dB/km attenuation coefficient and 16.7 ps/nm/km dispersion coefficient, an EDFA with NF of 5.5 dB, and a tunable optical filter with 3.5th-order super-Gaussian transfer function. We set the 3 dB bandwidth of the optical filter with values ranging between 137.5 GHz and 200 GHz. To study the performance of the proposed solution on higher-order signal modulation formats, we also considered a quadrature amplitude modulation (QAM) format with 16 constellation points (i.e., 16QAM). In addition, by varying the number of loops, we simulated a different number of optical spans. To simulate a B2B scenario, we removed the loop, maintaining only a single optical filter. Moreover, before receiving the signal, we placed a dispersion-compensating optical fiber of length equal to the loop fiber length to fully compensate for the chromatic dispersion effect. Then, we detected each of the four subchannels at the RX side by tuning the frequency of four coherent RX local oscillators. In [8], the optimum value for the RX low-pass electric filter 3 dB bandwidth was found to be equal to half of the symbol rate. Accordingly, we considered 3 dB bandwidth values for the RX filters equal to 16 GHz. Once detected, we collected the SNR of each subchannel and sent all of them to the optimization algorithm. The algorithm probed several times the VPI setup to obtain the measurements needed to derive the gradient, calculated a new set of frequencies, and fed that to VPI, triggering a new iteration of the simulation setup / iterative optimization process.

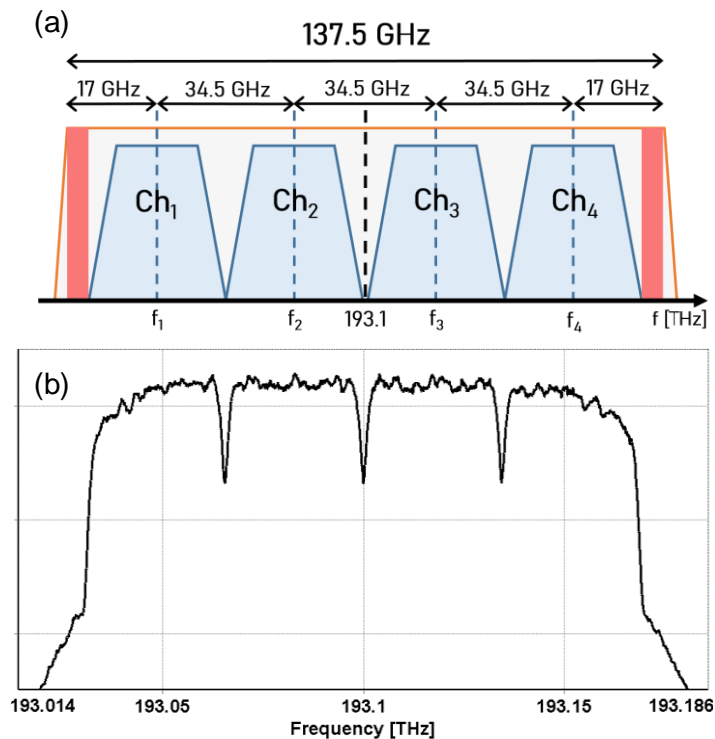

Fig. 4. (a) Schematic view of the considered default superchannel configuration with the four equally spaced subchannels. (b) Spectrum of the VPI-generated superchannel for the default case, i.e., a = 0.1, number of loops = 2, $BW_{SC}$ = 137.5 GHz.

In addition to the four-subchannel superchannel scenario, we also extended our simulations up to ten subchannels; the results for such configurations are reported at the end of this section. For the four-subchannel scenario discussed above, we considered nine transmission cases corresponding to different parameter configurations. Table 1 summarizes these nine cases. In particular, we considered a default case with $BW_{SC}$ equal to 137.5 GHz, roll-off factor a, for all the four transmitted signals, equal to 0.1, and two loops (i.e., twice the cascade of an 80 km length SSMF, an EDFA, and a superchannel filter). In addition, we assumed the subchannels to be equally spaced, i.e., $D_{eq-dis}$ = [17, 34.5, 34.5, 34.5, 17] GHz. We depict a schematic representation of such default configuration in Fig. 5a, whereas, in Fig. 5b, we show the spectrum as seen before the reception phase in VPI. Considering such default parameters, the received four subchannel SNR values were equal to $SNR_{SC}(D_{eq-dis})$ = [13.55, 16.86, 16.91, 13.15] dB. This resulted in an average superchannel SNR value of 15.12 dB, with 13.15 dB being the minimum among the four subchannel SNR values. Once integrated with the two optimization algorithms presented in Section 2, we found the optimized sets of distances $D^*_{obj1}$ and $D^*_{obj2}$. In particular, the set $D^*_{obj1}$ = [18.25, 33.25, 33.25, 33, 19.75] GHz corresponds to the solution that maximized the superchannel average SNR, with a step size of 0.25 GHz. The superchannel average SNR value associated with such configuration was 15.48 dB, yielding a 0.36 dB improvement with respect to the equally spaced initial distances $D_{eq-dis}$. In addition, again employing a frequency step equal to 0.25 GHz, we obtained the set $D^*_{obj2}$ = [19, 33, 31.75, 32.75, 21] GHz for the maximization of the minimum subchannel SNR and its corresponding set of SNR values $SNR^*_{obj2}(D^*_{obj2})$ = [14.34, 14.58, 14.37, 14.38] dB. We see that the algorithm flattened the SNR values of all the subchannels. This is expected in such an optimization objective, where the algorithm iteratively tries to improve the minimum, eventually bringing all the subchannels to the same level. The optimal minimum SNR value was found to be 14.34 dB. Such results translated in a 1.19 dB improvement with respect to equally spaced initial distances $D_{eq-dis}$. The algorithm iterations needed to obtain the previous solutions were 23 for the superchannel average SNR maximization problem and 9 for the minimum SNR maximization. Note that the number of iterations is important because it affects (linearly) the number of probes and monitoring calls.

Table 1. Simulation Parameters of the Nine Considered Cases

| # Case | Mod. Format | Roll-off | # Spans | # Filter | BW [GHz] |
|---|---|---|---|---|---|
| 1(DEF.) | QPSK | 0.1 | 2 | 2 | 137.5 |
| 2 | 16QAM | 0.1 | 2 | 2 | 137.5 |
| 3 | QPSK | 0.1 | 2 | 2 | 150 |
| 4 | QPSK | 0.15 | 2 | 2 | 137.5 |
| 5 | QPSK | 0.1 | 10 | 5 | 137.5 |
| 6 (B2B) | QPSK | 0.1 | 0 | 1 | 137.5 |
| 7 (B2B) | QPSK | 0.1 | 0 | 1 | 200 |
| 8 (B2B) | QPSK | 0.15 | 0 | 1 | 137.5 |
| 9 (B2B) | QPSK | 0.15 | 0 | 1 | 200 |

The evolution of the two optimization objectives as a function of the iteration number for the default case and equidistant subchannels as a starting point is shown in Fig. 6. We observe that the stochastic subgradient method is not a descent method (like gradient descent). Note that in this figure, we consider the negative of a convex/descent problem; we search for the maximum, so the objective ascends. Also, note that the number of iterations depends on the considered starting point. An overview of the default case-related results is shown in Fig. 7, where we graphically represented the SNR values obtained by the four subchannels considering the equidistant initial conditions and after the completion of the optimization algorithm for different step sizes.

The results related to the nine considered cases of Table 1 are summarized respectively in Table 2 for the average SNR maximization algorithm and Table 3 for the minimum subchannel SNR maximization. With respect to the default case (case 1), in cases 2 to 5, we considered variations of the: modulation format (case 2), superchannel filter bandwidth (case 3), roll-off factor (case 4), and span number (case 5). Regarding this last configuration, increasing the number of spans also implied the presence of nonlinear interference and other wavelength-dependent impairments, such as the gain ripple. In this case, employing the proposed algorithm, we improved Obj#1 and Obj#2 of 0.50 dB and 1.91 dB, respectively, with respect to the scenario with equally distanced subchannels. More in general, the obtained results reported in Tables 2 and 3 show how, independently from the variation of the aforementioned parameters, the proposed optimization solution was always able to improve the considered objectives. In addition, as per the default case, we ran simulations considering ten different random starting frequencies for cases 2, 3, 4, and 5. In all such cases, the algorithm always converged on the same optimal solution.

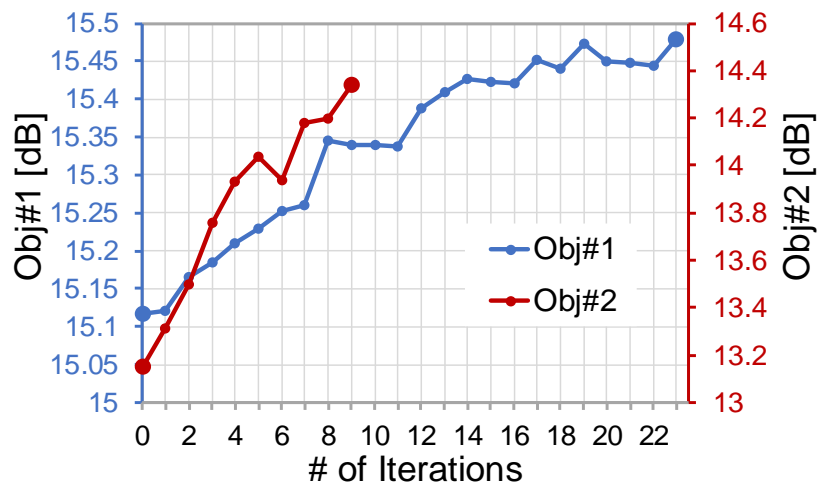

Fig. 5. Evolution of the SNR related to the two optimization problems as a function of the iteration number for the default case.

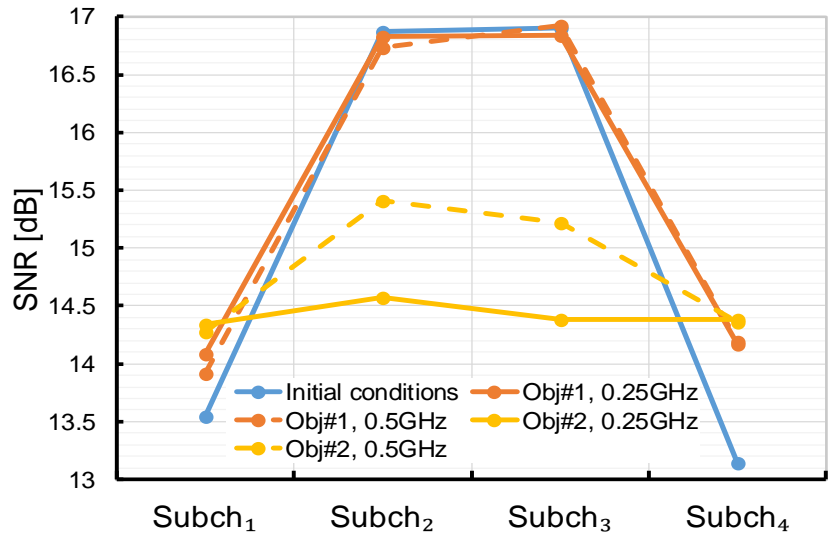

Fig. 6. SNR values of the four subchannels for the default case. The blue line represents the SNR values related to the initial conditions (i.e., equally spaced subchannels), whereas the brown and the yellow lines represent the SNR values after Obj#1 and Obj#2 optimization, respectively. For both the objectives, the solid lines refer to a step size of 0.25 GHz and the dashed lines to a step size of 0.5 GHz.

The cases from 6 to 9 represent four different B2B scenarios. We considered them to understand the contribution of the filtering and the crosstalk effects. As shown in Fig. 2, when the distances between subchannels range between 34.5 ±2 GHz, the crosstalk from the two adjacent subchannels is negligible. In cases 7 and 9, where we set a very broad filter 3 dB bandwidth (i.e., 200 GHz), almost no improvement was observed after the optimization with respect to the equidistant superchannel configuration. This happens because our starting conditions (i.e., 34.5 GHz distanced subchannels) already guaranteed an impairment-free transmission, and the optimization increases further all the distances, but with negligible improvement. On the other hand, when considering B2B and tighter filter bandwidth (i.e., cases 6 and 8), crosstalk- and filter-related penalties appears, and therefore our solution can improve the considered objectives. As expected, case 8, where we considered a higher roll-off factor, has a lower initial (average) SNR value than case 6. However, its improvement with respect to the equidistant configuration is relatively lower since both crosstalk and filtering effects are strong, and not enough optimization space is present.

Table 2. Optimization Results for the Superchannel Average SNR Value

| Case | Average SNR eq. distanced subch. [dB] | Step size [GHz] | # Iterations | Average SNR (Obj#1) [dB] | Brute-force ±2 GHz around opt. solution | | Improvement wrt worst case [dB] | Improvement wrt eq. distanced subch. [dB] |
|---|---|---|---|---|---|---|---|---|
| | | | | | Best average SNR [dB] | Worst average SNR [dB] | | |
| 1 | 15.12 | 0.25 | 23 | 15.48 | 15.48 | 13.01 | 2.47 | 0.36 |
| | | 0.50 | 7 | 15.44 | | | | |
| 2 | 15.65 | 0.25 | 17 | 16.13 | 16.13 | 12.10 | 4.03 | 0.48 |
| | | 0.50 | 10 | 16.06 | | | | |
| 3 | 15.40 | 0.25 | 43 | 16.15 | 16.28 | 13.56 | 2.72 | 0.88 |
| | | 0.50 | 24 | 16.28 | | | | |
| 4 | 15.30 | 0.25 | 9 | 15.48 | 15.48 | 12.71 | 2.77 | 0.18 |
| | | 0.50 | 12 | 15.46 | | | | |
| 5 | 8.56 | 0.25 | 24 | 9.06 | 9.06 | 7.79 | 1.27 | 0.50 |
| | | 0.50 | 22 | 9.06 | | | | |
| 6 | 21.25 | 0.25 | 31 | 21.62 | 21.62 | 15.55 | 6.07 | 0.37 |
| | | 0.50 | 16 | 21.33 | | | | |
| 7 | 23.29 | 0.25 | 36 | 23.33 | 23.33 | 16.08 | 7.25 | 0.04 |
| | | 0.50 | 19 | 23.33 | | | | |
| 8 | 20.92 | 0.25 | 47 | 20.99 | 21.04 | 16.12 | 4.92 | 0.12 |
| | | 0.50 | 21 | 21.04 | | | | |
| 9 | 22.86 | 0.25 | 43 | 22.83 | 22.86 | 18.87 | 3.96 | 0.00 |
| | | 0.50 | 21 | 22.81 | | | | |

In order to provide a benchmark for the obtained results, in Tables 2 and 3, we also report the best and the worst SNR values obtained employing a brute-force approach. In particular, we considered a ±2 GHz interval around the optimal solutions found by the algorithm (i.e., **D**$^*_{obj1}$ and **D**$^*_{obj2}$) and collected the SNR values related to all the possible distance combinations in such interval, with a frequency step size of $f_{step}^{BF}$ = 1 GHz. For each considered case listed in Tables 2 and 3, the total number of brute-force iterations was 5N. A smaller step, e.g., $f_{step}^{BF}$ = 0.25 GHz, would result in an extremely large number of scenarios and prohibitive simulation time. In a real network scenario, the worst SNR value found in this frequency range can represent the result of a soft failure occurring while the network is operating (e.g., ±2 GHz drift of each of the lasers). This worst-case should be considered by planning a high margin for the SNR and selecting an adequate modulation format. Instead, with our solution, we reduce this need. In particular, regarding the default case, with respect to such a soft failure scenario, we achieved improvements of 2.47 dB and 3.73 dB for the two considered objectives, respectively. Note that the proposed method would run periodically or on demand when e.g. we observe through monitoring that the performance (SNR or BER) has fallen below a specified threshold. Thus, the proposed method requires a margin, which can however be set to e.g. 1 dB from the optimum, and still achieve considerable savings of 1.47 dB and 2.73 dB, respectively. From Tables 2 and 3, we can see that the proposed method reached almost always the optimal SNR values identified by the brute-force approach. However, in case this did not happen (i.e., case 9 in Table 2), the optimal found solution almost equalized the best one. In particular, the solution related to case 9 missed the best one by 0.03 dB. Note that such a brute-force approach is unsuitable to be used in a real network, mainly for two reasons. On the one hand, because it has an execution time that is exponential to the subchannel number. In fact, considering an explored spectrum range $\Delta F$ = 4 GHz (i.e., ±2 GHz) and a $f_{step}^{BF}$= 1 GHz, we would have to configure and monitor the network $[1 + (\Delta F/ f_{step}^{BF})]^N$ times. Assuming a monitoring time of 5 minutes, this would take more than two days to complete, even for four subchannels. This increases even further for finer granularity. On the other hand, because it operates blindly, that is, without considering the subchannel SNR values and thus potentially making infeasible some of the connections. Comparing that with a few tens of monitoring calls (i.e., $I \cdot M$, with $M$ = 2) used by the proposed solution gives a clear benefit of employing our approach instead of the brute-force one.

Table 3. Optimization Results for the Minimum Superchannel Subchannel SNR Value

| Case | Min SNR eq. distanced subch. [dB] | Step size [GHz] | # Iterations | Min SNR (Obj#2) [dB] | Brute-force ±2 GHz around opt. solution | | Improvement wrt worst case [dB] | Improvement wrt eq. distanced subch. [dB] |
|---|---|---|---|---|---|---|---|---|
| | | | | | Best min SNR [dB] | Worst min SNR [dB] | | |
| 1 | 13.15 | 0.25 | 9 | 14.34 | 14.34 | 10.61 | 3.73 | 1.19 |
| | | 0.50 | 4 | 14.28 | | | | |
| 2 | 13.63 | 0.25 | 8 | 14.86 | 14.86 | 10.03 | 4.83 | 1.23 |
| | | 0.50 | 4 | 14.82 | | | | |
| 3 | 13.74 | 0.25 | 23 | 15.76 | 15.78 | 10.61 | 5.17 | 2.04 |
| | | 0.50 | 10 | 15.78 | | | | |
| 4 | 13.21 | 0.25 | 8 | 14.17 | 14.18 | 10.50 | 3.68 | 0.97 |
| | | 0.50 | 3 | 14.18 | | | | |
| 5 | 6.19 | 0.25 | 17 | 8.10 | 8.10 | 6.60 | 1.50 | 1.91 |
| | | 0.50 | 6 | 8.00 | | | | |
| 6 | 18.97 | 0.25 | 10 | 20.08 | 20.08 | 12.84 | 7.24 | 1.11 |
| | | 0.50 | 5 | 20.03 | | | | |
| 7 | 23.04 | 0.25 | 26 | 23.13 | 23.19 | 15.89 | 7.30 | 0.15 |
| | | 0.50 | 16 | 23.19 | | | | |
| 8 | 18.68 | 0.25 | 10 | 19.41 | 19.41 | 13.58 | 5.83 | 0.73 |
| | | 0.50 | 5 | 19.30 | | | | |
| 9 | 22.55 | 0.25 | 26 | 22.58 | 22.66 | 18.64 | 4.02 | 0.11 |
| | | 0.50 | 24 | 22.66 | | | | |

To further assess the effectiveness of the proposed method, we also integrated the optimization algorithm in scenarios simulated with six, eight, and ten subchannels. Apart from the TXs and RXs number, the rest of the setup was exactly as that depicted in Fig. 4. For all subchannel scenarios (i.e., six, eight and ten subchannels), we considered only one transmission case, similar to the default one of the four-subchannel configuration (i.e., case 1 in Table 1). In particular, for the scenario with six subchannels ($N$ = 6), we transmitted six 32 GBd QPSK equally spaced signals, with roll-off factor equal to 0.1, in a loop of two spans, filter 3 dB bandwidth of 200 GHz, and a minibatch number $M$ = 3. Similarly, we considered the same setup for the scenarios with eight and ten subchannels (i.e., $N$ = 8 and $N$ = 10, respectively), but a filter 3 dB bandwidth equal to 275 GHz and 340 GHz, respectively, and

$M = N/2$. Note that we used a frequency step size equal to 0.5 GHz for all the considered configurations and objectives.

For both the objectives, the achieved results always improved the figure of merit related to the starting conditions. In particular, concerning Obj#1, we improved the superchannel average SNR value, with respect to the equally spaced scenario, of 0.29 dB, 0.18 dB, and 0.06 dB, for the cases of six, eight, and ten subchannels, respectively. Moreover, for Obj#2, such improvements were equal to 1.67 dB, 1.33 dB, and 1.19 dB, respectively. We summarize the main results of the simulations related to Obj#2 in Table 4, where for the sake of comparison, we also report the corresponding case with four subchannels. As per the previous results, also in this case, it is worth reminding that the number of iterations required to reach the optimal solution depends on the considered subchannel starting positions. For instance, referring to Table 4 results, we can see that the iterations needed to find Obj#2 optimum in the case of six subchannels were lower compared to those needed to reach the optimum with eight subchannels. In addition, given the high amount of considered subchannels $N$, we ran the brute-force approach around the optimal found solution only within an interval of ±1 GHz. This resulted in a total number of brute-force iterations equal to $3N$. Such reduction was deemed necessary to complete the simulations in a reasonable time (i.e., within hours). This is also a further indication of the infeasibility of the brute-force approach in real scenarios. For Obj#1 and the three considered configurations (i.e., six, eight, and ten subchannels), when comparing our results to those related to a soft failure condition (i.e., the worst-case found within the brute-force interval), we achieved improvements equal to 2.79 dB, 2.60 dB, and 1.82 dB, respectively. To apply the proposed method we would need e.g. 1 dB of margin, still yielding quite high savings. Furthermore, as per the previous four-subchannel scenario, also in these cases, our optimal found solutions almost always matched the brute-force found optima. In particular, when this did not happen, we had the following accuracies: in scenarios with six and eight subchannels, for Obj#1, our optimal solutions were 0.01 dB and 0.02 dB away from the brute-force found optima, respectively. Instead, in the ten subchannels case, for Obj#2, our solution was 0.14 dB far from the optimum found with the brute-force approach.

Table 4. Results Comparison Between Different Subchannel Configurations for Obj#2

| Case | Min SNR eq. distanced subch. [dB] | Step size [GHz] | # Iterations | Min. SNR (Obj#2) [dB] | Brute-force ±1 GHz around opt. solution | | Improvement wrt worst case [dB] | Improvement wrt eq. distanced subch. [dB] |
|---|---|---|---|---|---|---|---|---|
| | | | | | Best Min. SNR [dB] | Worst Min. SNR [dB] | | |
| 4 subch., 137.5 GHz, 2 spans, 0.1 r.o. | 13.15 | 0.50 | 4 | 14.28 | 14.34 | 10.61 | 3.67 | 1.13 |
| 6 subch., 200 GHz, 2 spans, 0.1 r.o. | 11.90 | 0.50 | 16 | 13.57 | 13.57 | 9.53 | 4.04 | 1.67 |
| 8 subch., 275 GHz, 2 spans, 0.1 r.o. | 13.07 | 0.50 | 11 | 14.40 | 14.40 | 9.84 | 4.56 | 1.33 |
| 10 subch., 340 GHz, 2 spans, 0.1 r.o. | 13.11 | 0.50 | 14 | 14.30 | 14.44 | 9.81 | 4.49 | 1.19 |

## 5. Conclusions

Typically, superchannels are established with equidistant subchannels and a margin to account for laser drifts, filter misalignments, and wavelength-dependent effects. Leveraging the network feedback and taking into account the current network conditions can sense and correct the above issues, trace the real optimum, reduce the margin, and increase the efficiency of the superchannel. We proposed a closed control loop process for the spectral optimization of the superchannel subchannels. In particular, we observed that the problems of maximizing the superchannel average SNR and the minimum subchannel SNR values are concave near the operation point. We developed a solution that probes and monitors the network and then uses the stochastic subgradient method to optimize the chosen objective. The proposed stochastic subgradient method was chosen to be robust to monitoring errors and noise. We implemented and verified the proposed process by means of co-simulations in VPI and MATLAB.

The obtained results showed excellent performance for all the considered configurations. In particular, we considered superchannels composed of four, six, eight, and ten subchannels, with uniform characteristics, such as symbol rate and roll-off. We compared the optimal found solutions with the optima retrieved using a brute-force approach. We were able in all the considered scenarios to approach such optima. Moreover, with respect to the equidistant case, we improved for a uniform four-subchannel superchannel with 32 GBd symbol rate, 0.1 roll-off factor, and a superfilter bandwidth of 137.5 GHz, the superchannel average SNR and the minimum subchannel SNR values of 0.36 dB and 1.19 dB, respectively. Furthermore, considering a soft failure of ±2 GHz subchannel frequency drifts around the optimum, we would need a 2.47 dB and 3.73 dB margin, respectively. Applying the proposed method would reduce such margin to e.g. 1 dB, yielding quite high savings.

**Acknowledgments**

This work was partially funded by the ONFIRE project Horizon 2020 Framework Programme (765275).